\documentclass[twoside]{article}
\usepackage{fleqn,espcrc2,epsfig}

\usepackage{graphicx}
\usepackage[figuresright]{rotating}

\newcommand{\AmS}{{\protect\the\textfont2
  A\kern-.1667em\lower.5ex\hbox{M}\kern-.125emS}}
\def\MON{MONOLITH}
\def\numu{\nu_\mu}

\title{\MON: a next generation experiment for atmospheric neutrinos}

\author{P. Antonioli \address{Istituto Nazionale di Fisica Nucleare, 
        Sezione di Bologna, \\ 
        Via Irnerio 46, 40126 Bologna, Italy \\
        Pietro.Antonioli@bo.infn.it}%
        \thanks{on behalf of the \MON ~Collaboration}}
\begin{document}

\begin{abstract}
\MON ~is a massive magnetized tracking calorimeter, optimized for the detection
of atmospheric muon neutrinos, proposed at the Gran Sasso laboratory
in Italy.
The main goal is to establish (or reject) the neutrino oscillation
hypothesis through an explicit observation of the full first oscillation
swing (the ``$L/E$ pattern''). Its performance, status and prospects
are briefly reviewed.
\vspace{1pc}
\end{abstract}

\maketitle

\section{INTRODUCTION}
Although the evidence for neutrino oscillation is growing year after
year, the final proof that the observed anomalies are due to
oscillation is still outstanding.
Alternative models (neutrino decay, potential decoherence effect, 
influence of large extra-dimensions) to explain the Super-Kamiokande
(SK) atmospheric neutrinos results
have been proposed. On the experimental side, as pointed out also 
in recent Neutrino 2000 conclusions ``{\it there is as yet no 
direct evidence of an oscillation pattern}'' \cite{Ellis} and
the oscillation para\-meters in the atmospheric sector 
are still measured with a poor precision.

The \MON ~experiment has been designed exactly to reach
these physics goals, that is prove $\nu$ oscillations through the
observation of the oscillation pattern and improve the measurements
of parameters affecting the oscillations (if oscillation is the
pysical case). The experimental method that is proposed is designed 
to overcome the li\-mits of current atmospheric neutrino experiments through
a systematic free and Monte Carlo independent measurements of
the parameters.
\MON ~will also test the (now disfavored) hypothesis of mixing with 
sterile neutrinos searching for matter induced effects. Further
items of research, not discussed here, are the use of the 
huge sample of events coming from the CNGS beam to complement 
the atmospheric neutrino sample, the measurement of the cosmic muon
spectrum at TeV energies and the use of the detector as
a muon charge-ID detector for a beam coming from a $\nu$-factory.

\begin{table}[phb]
\label{table:detparm}
\vspace{-0.5cm}
\begin{tabular}{ll}
\hline \hline
Mass & 35 kt \\ 
Magnetic Field & 1.3 T \\
Dimensions       & (30.0$\times$14.5) m$^2$$\times$13.1 m\\
Space resolution &  1 cm \\
Time resolution  &  1 ns \\
GSC counters     &  54000 m$^2$ \\
External Veto    &  1500 m$^2$ \\
$p_{\mu}$ resolution &  $\approx$ 8\% (FC),
$\approx$ 20\% (PC)\\
$E_{h}$ resolution & 90\%$/\sqrt{E~\rm{(GeV)}} \oplus 30\%$ \\    
\hline \hline
\end{tabular}\\[2pt]
\caption{\MON ~main parameters. FC: fully contained events, PC:
partially contained. The External Veto consists of scintillator
counters.}
\end{table}

\section{APPARATUS AND EXPERIMENTAL APPROACH}
To explicitly detect an oscillation pattern in the $L/E$ spectrum
of atmospheric muon neutrinos, the energy $E$ and direction
$\theta$ of the incoming neutrino have to be measured in each
event. The latter can be estimated from the direction $\theta _{\mu}$ of the 
muon produced from the $\numu$
charged-current interaction. The neutrino energy $E$ can be
obtained by means of  energy  measurements of the muon $E_{\mu}$
and of the hadrons produced in the interaction ($E_{h}$). In order to make the
oscillation pattern detectable, the $L/E$ ratio 
has to be measured with a FWHM error
smaller than half of the modulation period.
The resolution on $L/E$ improves at high energies, mostly because the
muon direction gives a better estimate of the neutrino
direction. Therefore the ability to measure high momentum muons (in the
multi-GeV range) in partially contained events, which is rather 
limited in the on-going atmospheric neutrino experiments, 
is particularly rewarding.

A complete description of the detector structure can be found in 
\cite{Proposal}. The apparatus consists of two modules,
a stack of 125 horizontal 8 cm thick iron plates each with a surface
of 15.0$\times$14.5 m$^2$, interleaved with 2.2 cm gaps housing the
sensitive elements. The main detector parameters and resolutions 
are summarized in Table 1.
Taking into
account needs of such a large area detector and fast mass production,
Glass Spark Counters \cite{Proposal}, derived from resistive-plate
chambers, have been chosen as active elements.
The magnetization of iron plates is obtained through a vertical
slot crossing all the stack: 
each field line 
lays entirely inside an
individual plane. 

\begin{figure}[pb]
\vspace{-2 cm}
\label{figure:accmag}
\vspace{0.5cm}
\mbox{\epsfig{file=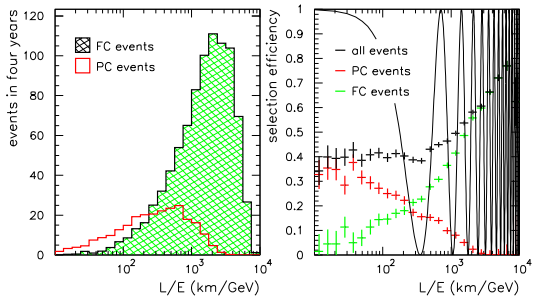, width=\linewidth}}
\vspace{-1cm}
\caption{Selected events (\textit{left}) and
overall selection efficiency and selection efficiency
for FC and PC contained events as a function of $L/E$ (\textit{right}).
The $\nu_{\mu}$ survival probability using present best-fit value
of SK is superimposed.}
\end{figure}

To avoid systematic effects and Monte Carlo dependencies,
we use the ratio between the measured down-going 
neutrino flux as a \textit{near} unoscillated reference flux and the upward
neutrinos flux as a \textit{far} oscillated one. Thus we associate
for each downward neutrino its \textit{mirror} pathlength, as
originally proposed in \cite{Pic97}. This method is to a large extent
insensitive to uncertainties in the knowledge of atmospheric
neutrino fluxes and neutrino cross sections as well detector inefficiencies.
It is based on the up/down symmetry of the neutrino fluxes, which holds
at typical \MON ~energies ($E_{th}>1.5$ GeV). In this framework, the
atmospheric neutrinos represent an ideal case for a disappearance
experiment.

\section{DETECTION OF THE $L/E$ PATTERN}
The event selection has been tuned to obtain the $L/E$
resolution required to detect the pattern. Muons are reconstructed or
via energy loss in the iron or via magnetic analysis.
Minimal requirements have
been applied to the reconstructed muon momentum ($p_{\mu}>1.5$ GeV) and, 
for outgoing muons, to the track-length $l$ ($l>4$ m). Additional
cuts (for details see \cite{Proposal}) on combinations of
the observables $E_{\mu}$, $E_h$ and $\theta _{\mu}$ are applied
to ensure the required $L/E$ re\-solution and on visible vertex
coordinates, muon direction, external veto information to
reject the cosmic muon background. With these cuts, we expect in 4
years of data taking (see Fig. 1) to select for phy\-sical analysis 
$\approx$ 1200 events (80\% fully contained and 20\% partially contained) of 
down-going $\numu$ (in case of no oscillation this number is 
the same for up-going).

\begin{figure}[h]
\label{figure:pattern}
\vspace{-0.5 cm}
\begin{center} 
\mbox{\epsfig{file=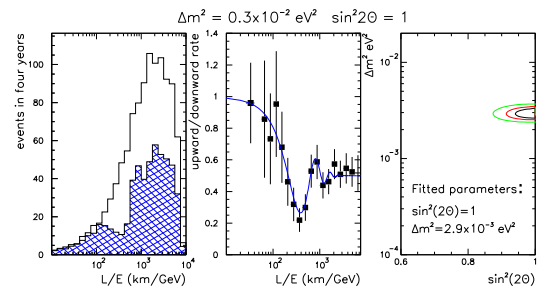, width=\linewidth}}
\end{center}
\vspace{-1cm}
\caption{The $L/E$ spectrum of upward muon neutrino events (hatched
area) and the $L/E$ ``mirrored'' spectrum of downward muon neutrino
events (open area) (\textit{left}); their ratio with our 
best-fit superimposed (\textit{center}); results of the fit (68\%,
90\% and 99\% C.L. shown) (\textit{right}).
}
\end{figure}

The gain in acceptance at small values of $L/E$ due to the magnetic field
analysis is noteworthy: as shown in Fig.~1 the inclusion
of partially contained events increases the \MON ~selection efficiency
by more than a factor two in the region where the first semi-period of
the oscillation is expected according to the SK results.
The cost of the magnetization, when compared to the whole cost, 
is around 8\%, but, in practice, has an effect similar
to that of doubling the total mass.
 
In Fig.~2 we show, for a particular choice of oscillation
parameters, how \MON ~detects the oscillation pattern using the 
ratio between up-going and down-going events. 
Note that around $\Delta m^2\approx10^{-3}$eV$^2$ \MON ~can
measure the $\Delta m^2$ value with 6\% precision.

An accurate analysis \cite{Proposal} of the capability of \MON ~to confirm
the oscillation hypothesis with respect to other alternative models
has been performed. Around
current SK values, \MON ~is able to reject the decay hypothesis
at 95\% C.L. or better with 99.5\% efficiency. In Fig. 3
the expected allowed regions of the $\nu _{\mu}\rightarrow\nu _{\tau}$
oscillation parameters after 4 years of data taking with \MON ~ are
shown for four different values of $\Delta m^2$. It is worthwhile
to note that the progress in the knowledge of the $\Delta m^2$
expected by \MON ~with respect to SK, is
quantitatively similar to the one done by SK with
respect to Kamiokande.

\section{STATUS AND PROSPECTS}
The \MON ~Collaboration counts now 85 physicists and 17 institutions.
After preliminary detector studies \cite{LoI,Progress},
a Proposal \cite{Proposal} has been recently submitted 
to the LNGSC. 
Cosmic ray and beam tests have been carried out
during the last two years at LNGS and CERN/PS to study up/down
discrimination and hadronic e\-nergy resolution. 
Assuming approval in early 2001, the first
module is foreseen to be operational by the end of 2004 and the
detector to be completed by the end of 2006.

The superior L/E resolution of the detector will allow detection of
the first oscillation period. \MON ~will thus prove (or disprove)
the oscillation hypothesis. It will also highly improve the measurement of
the oscillation parameters. These qualitative and quantitative
improvements qualify \MON ~as a next generation neutrino oscillation
detector.

We finally note that the basic parameters
of detectors that are proposed for neutrino beams from muon storage rings
look very similar to those of \MON ~(so that the expected
performances should roughly apply to \MON). The useful life of
the detector might be extended accordingly.

\textbf{Acknowledgements.} I wish to address my warmest thanks
to Stefano Ragazzi and Francesco Terranova who provided me useful
material to prepare this talk and Rosario Nania for
helpful discussions.

\begin{figure}[bp]
\label{figure:sensitivity}
\vspace{-1.cm} 
\mbox{\epsfig{file=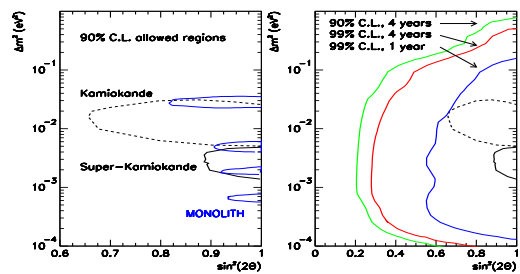, width=\linewidth,height=4cm
}}
\vspace{-1.5cm}
\caption{ \textit{Left:} expected allowed regions of 
$\nu_\mu-\nu_\tau$ oscillation
parameters for MONOLITH after four years of exposure: the results of
the simulation for $\Delta m^2 = 0.7, 2, 5, 30 \times 10^{-3}$ eV$^2$
and maximal mixing are shown \textit{Right:} exclusion curve assu\-ming
no oscillation.} 
\end{figure}

\end{document}